\documentclass[12pt]{article}
\usepackage{mathrsfs}
\usepackage{amsmath,amssymb,latexsym,color,cancel,graphicx,bbm,colortbl}
\usepackage[english]{babel}
\usepackage[latin1]{inputenc}
\usepackage{ragged2e}
\begin{document}
\date{}

\title{$SU(1,1)$ and $SU(2)$ Perelomov number coherent states: algebraic approach for general systems}
\author{D. Ojeda-Guillén,$^{a}$\footnote{{\it E-mail address:} dojedag@ipn.mx}\quad M. Salazar-Ramírez,$^{a}$\\ R.D. Mota$^{b}$ and V.D. Granados$^{c}$} \maketitle

\begin{minipage}{0.9\textwidth}

\small $^{a}$ Escuela Superior de Cómputo,
Instituto Politécnico Nacional,  Av. Juan de Dios Bátiz esq. Av. Miguel Othón de Mendizábal, Col. Lindavista, Del. Gustavo A. Madero, C.P. 07738, México D. F., Mexico.\\

\small $^{b}$ Escuela Superior de Ingeniería Mecánica y Eléctrica, Unidad Culhuacán,
Instituto Politécnico Nacional,   Av. Santa Ana No. 1000, Col. San
Francisco Culhuacán, Delegación Coyoacán, C.P. 04430,  México D. F., Mexico.\\

\small $^{c}$ Escuela Superior de Física y Matemáticas,
Instituto Politécnico Nacional,
Ed. 9, Unidad Profesional Adolfo López Mateos, C.P. 07738 México D. F., Mexico.\\

\end{minipage}

\begin{abstract}

We study some properties of the $SU(1,1)$ Perelomov number coherent states. The Schrödinger's uncertainty relationship
is evaluated for a position and momentum-like operators (constructed from the Lie algebra generators) in these number coherent states. It is shown that this relationship is minimized for the standard coherent states. We obtain the time evolution of the number coherent states by supposing that the Hamiltonian is proportional to the third generator $K_0$ of the $su(1,1)$ Lie algebra. Analogous results for the $SU(2)$ Perelomov number coherent states are found. As examples, we compute the Perelomov coherent states for the pseudoharmonic oscillator and the two-dimensional isotropic harmonic oscillator.

\end{abstract}


\section{Introduction}
Erwin Schrödinger introduced coherent states in quantum
mechanics while he was looking for a system which possessed a
classical behavior \cite{Scrho}. The coherent states were reintroduced
in quantum optics by the works of Glauber \cite{Glau}, Klauder
\cite{Klau,Klau2} and  Sudarshan \cite{Sudar}. These states are related to
the Heisenberg-Weyl group. Harmonic oscillator coherent states are the
most classical states, since they minimize the Heisenberg uncertainty relationship.

The coherent states for the one-dimensional harmonic oscillator
were generalized by introducing the displaced
number states or number coherent states of the harmonic oscillator. Boiteux and Levelut
defined these states by applying the Weyl operator to any excited state
$|n\rangle$ and they called them semicoherent states \cite{BandL}. Later, Roy and Singh \cite{Roy},
Satyanarayana \cite{Saty}, and Oliveira, Kim, Night and Bu$\check{z}$ek \cite{Oli} gave a detailed study of
the properties of these states. A few years later, Nieto \cite{Nieto2}
derived the most general form of these states.

However, the Heisenberg-Weyl is not the only group for which we can
construct coherent states. In the 70's, the works of A. O. Barut and
L. Girardello \cite{BandG} and Perelomov \cite{Perel} generalized
the concept of coherent states to general systems related to any algebra of a
symmetry group. These approaches remain as current research fields as
it is shown in references \cite{Gazeaulibro,Klimov}. In particular,
related to the $su(2)$ and $su(1,1)$ Lie algebra several works have been published,
and some of them are in references \cite{Eberly,BrifHO,BrifHO2}.

On the other hand, the Heisenberg uncertainty relationship was
generalized by the work of Schrödinger \cite{Schro2} and Robertson
\cite{Rober} for any two observables. Recently, these uncertainty
relationships were generalized to several observables and several
states \cite{Trifo}. With these results, the harmonic oscillator
coherent states have been generalized too, by
constructing states that minimize those uncertainty relations. These
states which minimize uncertainty relationships have been widely studied
\cite{Trifo2,Trifo3,Trifo4} and are called intelligent states \cite{Ara}.

The Perelomov's coherent states were extended by Gerry, who studied the
$SU(1,1)$ number coherent states \cite{Berry}. Gerry defined these states as the action
of the displacement operator onto any $SU(1,1)$ excited state and obtained
a general form of these states in terms of the Bargmann V functions. Moreover,
he showed that these states are the eigenfunctions of the degenerate
parametric amplifier, by an appropriate choice of the coherent state
parameters.

Recently, two important applications of the $SU(1,1)$ and $SU(2)$ Perelomov number coherent states
have been founded. It has been shown that the number coherent states of the two-dimensional
harmonic oscillator are the eigenfunctions of the non-degenerate
parametric amplifier \cite{Nos1} and of two coupled oscillators \cite{Nos2}.

The aim of the present work is to study the dispersion and time evolution of the Perelomov number coherent states for
the $su(1,1)$ and $su(2)$ Lie algebras. We show that the only minimum uncertainty states are the standard coherent states, even if we consider their time evolution. Finally, we apply our results to construct the Perelomov number coherent states of the pseudoharmonic oscillator (related to the $su(1,1)$ Lie algebra) and the two-dimensional harmonic oscillator (related to the $su(2)$ Lie algebra).

This work is organized as follows. In Section $2$ we introduce the
Perelomov number coherent states for the $su(1,1)$ Lie algebra.
We obtain the expected values of the
Lie algebra generators in the Perelomov number coherent states.
We define two position and momentum-like operators for the $su(1,1)$ Lie
algebra and we prove that standard Perelomov coherent states are of minimum
uncertainty, accordingly to the Schrödinger's uncertainty relationship.
By supposing that the Hamiltonian is proportional to one of the
generators of the $su(1,1)$ Lie algebra, we obtain the time evolution of the
Perelomov number coherent states. All previous results are applied to compute
the Perelomov number coherent states of the pseudoharmonic oscillator.
In Section $3$, we obtain the analogous results of the previous section
for the $su(2)$ Lie algebra Perelomov number coherent states. For this group, we calculate
the $SU(2)$ coherent states of the two-dimensional harmonic oscillator. Finally, we give some concluding
remarks.

\section{$SU(1,1)$ Perelomov number coherent states}

The $su(1,1)$ Lie algebra is spanned by the generators $K_{+}$, $K_{-}$
and $K_{0}$, which satisfy the commutation relations \cite{Vourdas}
\begin{eqnarray}
[K_{0},K_{\pm}]=\pm K_{\pm},\quad\quad [K_{-},K_{+}]=2K_{0}.\label{com}
\end{eqnarray}
The action of these operators on the Fock space states
$\{|k,n\rangle, n=0,1,2,...\}$ is
\begin{equation}
K_{+}|k,n\rangle=\sqrt{(n+1)(2k+n)}|k,n+1\rangle,\label{k+n}
\end{equation}
\begin{equation}
K_{-}|k,n\rangle=\sqrt{n(2k+n-1)}|k,n-1\rangle,\label{k-n}
\end{equation}
\begin{equation}
K_{0}|k,n\rangle=(k+n)|k,n\rangle,\label{k0n}
\end{equation}
where $|k,0\rangle$ is the lowest normalized state. The Casimir
operator $K^{2}=K_{\pm}K_{\mp}-K_{0}(K_{0}\mp1)$ for any irreducible
representation satisfies $K^{2}=k(k-1)$. Thus, a representation of
$su(1,1)$ algebra is determined by the number $k$. For the purpose of the
present work we will restrict to the discrete series only, for which
$k>0$.

The standard Perelomov coherent states $|\zeta\rangle$ are
defined as \cite{Perellibro}
\begin{equation}
|\zeta\rangle=D(\xi)|k,0\rangle,\label{defPCS}
\end{equation}
where $D(\xi)=\exp(\xi K_{+}-\xi^{*}K_{-})$ is the displacement
operator and $\xi$ is a complex number. From the properties
$K^{\dag}_{+}=K_{-}$ and $K^{\dag}_{-}=K_{+}$ it can be shown that
the displacement operator possesses the property
\begin{equation}
D^{\dag}(\xi)=\exp(\xi^{*} K_{-}-\xi K_{+})=D(-\xi),
\end{equation}
and the so called normal form of the displacement operator is given by
\begin{equation}
D(\xi)=\exp(\zeta K_{+})\exp(\eta K_{0})\exp(-\zeta^*
K_{-})\label{normal},
\end{equation}
where $\xi=-\frac{1}{2}\tau e^{-i\varphi}$, $\zeta=-\tanh
(\frac{1}{2}\tau)e^{-i\varphi}$ and $\eta=-2\ln \cosh
|\xi|=\ln(1-|\zeta|^2)$ \cite{Gerry}. By using this normal form of the displacement
operator and equations (\ref{k+n})-(\ref{k0n}), the Perelomov coherent states are found to
be \cite{Perellibro}
\begin{equation}
|\zeta\rangle=(1-|\zeta|^2)^k\sum_{s=0}^\infty\sqrt{\frac{\Gamma(n+2k)}{s!\Gamma(2k)}}\zeta^s|k,s\rangle.\label{PCN}
\end{equation}

The Perelomov number coherent states are defined as the action of
the displacement operator $D(\xi)$ on any state $|k,n\rangle$,
instead to the lowest state $|k,0\rangle$ of the Fock space  \cite{Berry}. This is
the obvious generalization of equation (\ref{defPCS}). Thus the states
\begin{equation}
|\zeta,k,n\rangle=D(\xi)|k,n\rangle=\exp(\zeta K_{+})\exp(\eta
K_{3})\exp(-\zeta^* K_{-})|k,n\rangle\label{defPCNS}
\end{equation}
are the $SU(1,1)$ Perelomov number coherent states.
The last equality is due to the normal form of the displacement
operator of equation (\ref{normal}). The Perelomov number coherent states in the Fock space are \cite{Nos1}
\begin{eqnarray}
|\zeta,k,n\rangle &=&\sum_{s=0}^\infty\frac{\zeta^s}{s!}\sum_{j=0}^n\frac{(-\zeta^*)^j}{j!}e^{\eta(k+n-j)}
\frac{\sqrt{\Gamma(2k+n)\Gamma(2k+n-j+s)}}{\Gamma(2k+n-j)}\nonumber\\
&&\times\frac{\sqrt{\Gamma(n+1)\Gamma(n-j+s+1)}}{\Gamma(n-j+1)}|k,n-j+s\rangle.\label{PNCS}
\end{eqnarray}

These states generalize the Perelomov coherent states (\ref{PCN}), which are obtained
by setting $n=0$ in last equation.

By using the Baker-Campbell-Hausdorff identity
\begin{equation}
e^{-A}Be^A=B+\frac{1}{1!}[B,A]+\frac{1}{2!}[[B,A],A]+\frac{1}{3!}[[[B,A],A],A]+...,
\end{equation}
and equations (\ref{com}), we can find the similarity transformations
$D^{\dag}(\xi)K_{+}D(\xi)$, $D^{\dag}(\xi)K_{-}D(\xi)$ and
$D^{\dag}(\xi)K_{0}D(\xi)$ of the $su(1,1)$ Lie algebra generators.
These results are given by
\begin{equation}
D^{\dag}(\xi)K_{+}D(\xi)=\frac{\xi^{*}}{|\xi|}\alpha
K_{0}+\beta\left(K_{+}+\frac{\xi^{*}}{\xi}K_{-}\right)+K_{+},\label{simiK+}
\end{equation}
\begin{equation}
D^{\dag}(\xi)K_{-}D(\xi)=\frac{\xi}{|\xi|}\alpha
K_{0}+\beta\left(K_{-}+\frac{\xi}{\xi^{*}}K_{+}\right)+K_{-},\label{simiK-}
\end{equation}
\begin{equation}
D^{\dag}(\xi)K_{0}D(\xi)=(2\beta+1)
K_{0}+\frac{\alpha\xi}{2|\xi|}K_{+}+\frac{\alpha
\xi^*}{2|\xi|}K_{-},\label{simiK0}
\end{equation}
where $\alpha=\sinh(2|\xi|)$ and
$\beta=\frac{1}{2}\left[\cosh(2|\xi|)-1\right]$.

Moreover, the expected values of the group generators $K_{\pm}, K_0$ in the Perelomov number coherent states
can be easily computed by using the similarity transformations of equations (\ref{simiK+})-(\ref{simiK0}).
Thus,
\begin{equation}
\langle \zeta,k,n|K_{+}|\zeta,k,n\rangle=\frac{\xi^*}{|\xi|}\sinh(2|\xi|)(k+n),
\end{equation}
\begin{equation}
\langle \zeta,k,n|K_{-}|\zeta,k,n\rangle=\frac{\xi}{|\xi|}\sinh(2|\xi|)(k+n),
\end{equation}
\begin{equation}
\langle \zeta,k,n|K_{0}|\zeta,k,n\rangle=\cosh(2|\xi|)(k+n).
\end{equation}

\subsection{Schrödinger's uncertainty relationship}

From the $SU(1,1)$ group ladder operators $K_{+}$ and $K_{-}$, we define the operators $X$ and $Y$ as \cite{Chinos}
\begin{eqnarray}
X\equiv K_{+}+K_{-},\quad\quad Y\equiv i(K_{+}-K_{-}).
\end{eqnarray}
With these equations we can compute the quadratic deviations of the
operators $X$ and $Y$ for the Perelomov number coherent states
\begin{equation}
(\Delta
X)^{2}_{n}=\langle\zeta,k,n|X^{2}|\zeta,k,n\rangle-\langle\zeta,k,n|X|\zeta,k,n\rangle^{2},\label{ddx}
\end{equation}
\begin{equation}
(\Delta
Y)^{2}_{n}=\langle\zeta,k,n|Y^{2}|\zeta,k,n\rangle-\langle\zeta,k,n|Y|\zeta,k,n\rangle^{2}.\label{ddy}
\end{equation}
The definition of the Perelomov number coherent states
(\ref{defPCNS}) and the similarity transformations, equations
(\ref{simiK+}) and (\ref{simiK-}), lead us to obtain
\begin{equation}
\langle\zeta|X^{2}|\zeta\rangle_{n}=\alpha^{2}(k+n)^{2}\left(2+\frac{\xi^*}{\xi}+\frac{\xi}{\xi^*}\right)
+2(n^2+2kn+k)\left[\left(2+\frac{\xi^*}{\xi}+\frac{\xi}{\xi^*}\right)(\beta^2+\beta)+1\right],
\end{equation}
\begin{equation}
\langle\zeta|X|\zeta\rangle_{n}=\frac{\alpha (k+n)}{|\xi|}(\xi^*+\xi),
\end{equation}
and
\begin{equation}
\langle\zeta|Y^{2}|\zeta\rangle_{n}=\alpha^{2}(k+n)^{2}\left(2-\frac{\xi^*}{\xi}-\frac{\xi}{\xi^*}\right)
+2(n^2+2kn+k)\left[\left(2-\frac{\xi^*}{\xi}-\frac{\xi}{\xi^*}\right)(\beta^2+\beta)+1\right],
\end{equation}
\begin{equation}
\langle\zeta|Y|\zeta\rangle_{n}=\frac{i \alpha (k+n)}{|\xi|}(\xi^*-\xi).
\end{equation}
By substituting these results into equations (\ref{ddx}) and
(\ref{ddy}) we obtain the quadratic deviations of the $X$ and $Y$
operators
\begin{equation}
(\Delta
X)^{2}_{n}=2(n^2+2kn+k)\left[\left(2+\frac{\xi^*}{\xi}+\frac{\xi}{\xi^*}\right)(\beta^2+\beta)+1\right],\label{dxn}
\end{equation}
and
\begin{equation}
(\Delta
Y)^{2}_{n}=2(n^2+2kn+k)\left[\left(2-\frac{\xi^*}{\xi}-\frac{\xi}{\xi^*}\right)(\beta^2+\beta)+1\right].\label{dyn}
\end{equation}
Hence, the product of these quadratic deviations is
\begin{equation}
(\Delta X)^{2}_{n}(\Delta Y)^{2}_{n}=4(n^2+2kn+k)^2\left\{(\beta^2+\beta)^2\left[4-\left(\frac{\xi^*}{\xi}+\frac{\xi}{\xi^*}\right)^2\right]+4(\beta^2+\beta)+1\right\}.\label{prodxyn}
\end{equation}

The Schrödinger's uncertainty relationship states that
the product of the quadratic deviations of any two operators $X$ and
$Y$ satisfy \cite{Schro2}
\begin{equation}
(\Delta X)^{2}(\Delta Y)^{2}\geq\langle F
\rangle^{2}+\frac{1}{4}\langle C \rangle^{2},\label{schro}
\end{equation}
where, $\langle C\rangle\equiv-i\langle[X,Y]\rangle$, and $\langle F
\rangle\equiv\langle \frac{1}{2}\{X,Y\}+\langle X\rangle\langle
Y\rangle \rangle$ is the quantum correlation of the operators $X$
and $Y$.

If we use equation (\ref{defPCNS}) and the similarity transformation
method to calculate the expectation values $\langle F \rangle$ and
$\langle C \rangle$ in a Perelomov number coherent state, we obtain
\begin{equation}
\langle\zeta,k,n|F|\zeta,k,n\rangle_{n}=2i(n^2+2kn+k)(\beta^2+\beta)\left(\frac{\xi^{*}}{\xi}-
\frac{\xi}{\xi^{*}}\right),\label{fn}
\end{equation}
\begin{equation}
\langle\zeta,k,n|C|\zeta,k,n\rangle_{n}=4(k+n)(2\beta+1)\label{cn}.
\end{equation}
By substituting the results of equations (\ref{prodxyn}), (\ref{fn})
and (\ref{cn}) into equation (\ref{schro}), we conclude that the number coherent states
are not of minimum uncertainty, accordingly to the Schrödinger's uncertainty relationship. However, for the Perelomov
coherent states $(n=0)$ the equality in (\ref{schro}) holds.
Therefore, the only states which minimize the Schrödinger's
uncertainty relationship are those obtained by applying the
displacement operator $D(\xi)$ to the lowest normalized state. This
result is in full agreement to that previously reported in
\cite{Perellibro}.

The study of the uncertainty relations is a cornerstone in the study
of squeezing. In fact, the change of shape of the radial probability
distribution between the turning point of the harmonic oscillator
coherent states can be interpreted (at least in part) as squeezing \cite{gerryoscillator}.

\subsection{Time evolution of the $SU(1,1)$ Perelomov number coherent states}

The time evolution operator $U(t)$ for an arbitrary Hamiltonian $H$
is defined as $U(t)=e^{-iHt/\hbar}$ \cite{Cohen}. Notice that in many problems
the Hamiltonian is proportional to the group operator $K_{0}$ \cite{mann}.
Thus, without lose of generality, we can write the time evolution operator as
\begin{equation}
U(t)=e^{-i\gamma K_{0}t/\hbar}.
\end{equation}
With the previous definition, the BCH identity and equation
(\ref{com}), we can compute the time evolution of the $SU(1,1)$
group ladder operators $K_{\pm}$ with the similarity transformations
\begin{equation}
K_{+}(t)=U^{\dag}(t)K_{+}U(t)=K_{+}e^{i\gamma t/\hbar},\label{k+t}
\end{equation}
\begin{equation}
K_{-}(t)=U^{\dag}(t)K_{-}U(t)=K_{-}e^{-i\gamma t/\hbar}.\label{k-t}
\end{equation}
Notice that we can obtain the same results by using
the Heisenberg equations. Thus, from equation (\ref{defPCNS}), the
time evolution of the Perelomov number coherent states $|\zeta(t),k,n\rangle$ is given by
\begin{equation}
|\zeta(t),k,n\rangle=U(t)|\zeta,k,n\rangle=U(t)D(\xi)U^{\dag}(t)U(t)|k,n\rangle.
\end{equation}
From equation (\ref{k0n}), the time evolution of the state $|k,n\rangle$ is given by
\begin{equation}
U(t)|k,n\rangle=e^{-i\gamma (k+n)t/\hbar}|k,n\rangle.\label{uk,n}
\end{equation}
On the other hand, from equations (\ref{k+t}) and (\ref{k-t}) we find
\begin{equation}
U(t)D(\xi)U^{\dag}(t)=e^{\xi K_{+}(-t)-\xi^* K_{-}(-t)}=e^{\xi(-t) K_{+}-\xi(-t)^* K_{-}},
\end{equation}
where we have introduced the time dependent complex $\xi(t)=\xi e^{i\gamma t/\hbar}$. Thus, the time evolution of the
displacement operator $D(\xi)$ is due to the time evolution of the complex $\xi$. The time evolution of the
displacement operator in its normal form is given by
\begin{equation}
D\left(\xi(t)\right)=U^{\dag}(t)D(\xi)U(t)=U^{\dag}(t)e^{\zeta K_{+}}e^{\eta K_{0}}e^{-\zeta^* K_{-}}U(t).
\end{equation}
If we introduce the complex $\zeta(t)=\zeta e^{i\gamma
t/\hbar}$, we obtain the time-dependent
displacement operator $D(\xi)$
\begin{equation}
D\left(\xi(t)\right)=e^{\zeta(t) K_{+}}e^{\eta K_{0}}e^{-\zeta(t)^* K_{-}}.\label{dtn}
\end{equation}
With the previous results and the equations (\ref{uk,n}) and
(\ref{dtn}), we obtain that the time dependent Perelomov number coherent states are
\begin{equation}
|\zeta(t),k,n\rangle=e^{-i\gamma (k+n)t/\hbar}e^{\zeta(-t) K_{+}}e^{\eta K_{0}}e^{-\zeta(-t)^* K_{-}}|k,n\rangle.\label{PNCSt}
\end{equation}
Thus, the time evolution of the number coherent states for the $SU(1,1)$ group is obtained by adding the phase $e^{-i\gamma (k+n)t/\hbar}$ and substituting $\zeta\rightarrow\zeta(-t)$ and $\zeta^*\rightarrow\zeta(-t)^*$ into equation (\ref{PNCS}). The expression of equation (\ref{PNCSt}) generalizes the Perelomov coherent states, which are recovered by setting $t=0$ and $n=0$. The results of this section can be extended to the cases in which the Hamiltonian depends on a linear combination of the algebra generators, instead of just $K_{0}$.

\subsection{$SU(1,1)$ number coherent states for the Pseudoharmonic Oscillator}

The pseudoharmonic oscillator is described by the one-dimensional potential
\begin{equation}
V(x)=\frac{1}{2}m\omega^2 x^2+\frac{\hbar^2}{2m}\frac{\alpha}{x^2},
\end{equation}
where $m$, $\omega$ and $\alpha$ represent the mass of the particle, the frequency and the
strength of the external field, respectively.
The normalized wave functions for the pseudoharmonic oscillator are given by \cite{SHD}
\begin{equation}\label{EMOR}
\Phi^{s}_n(\rho)=N_n\rho^{s}e^{-\frac{\rho}{2}}L^{2s-\frac{1}{2}}_n(\rho),\quad\quad N_n=\sqrt{\frac{n!}{\Gamma(n+2s+1/2)}},
\end{equation}
where $\rho=x^2$. The $su(1,1)$ Lie algebra generators of the pseudoharmonic oscillator can be constructed by using the
recursion relations among the associated Laguerre functions \cite{SHD}. These operators explicitly are
\begin{equation}
K_-=-\rho \frac{\partial}{\partial \rho}+s+\hat{n}-\frac{\rho}{2}, \quad\quad K_0=\hat{n}+s+\frac{1}{4}\nonumber
\end{equation}
\begin{equation}
K_+=\rho \frac{\partial}{\partial \rho}+s+\hat{n}+\frac{1}{2}-\frac{\rho}{2}.
\end{equation}
The action of ladder operators on the pseudoharmonic oscillator wave functions is
\begin{align}
K_+|s,n\rangle &=\sqrt{(n+1)(n+2s+1/2)}|s,n+1\rangle,\\
K_-|s,n\rangle &=\sqrt{n(n+2s-1/2)}|s,n-1\rangle,\\
K_0|s,n\rangle &=\left(n+s+1/4\right)|s,n\rangle.
\end{align}
By comparing these results with equations (\ref{k+n})-(\ref{k0n}), we obtain that the relationship between the group numbers $k,n$ and the quantum numbers $s,n$ satisfies $k\rightarrow s+1/4$ and  $n\rightarrow n$. The Perelomov number coherent states of the pseudoharmonic oscillator $\Psi_{PO}$ are obtained by substituting the states of equation (\ref{EMOR}) into equation (\ref{PNCS}). Thus, by interchanging the order of summations and using the relationships between the group and quantum numbers we obtain
\begin{align}
\Psi_{PO}=&\langle\rho|\zeta,k,n\rangle=\left(1-|\zeta|^2\right)^{s+n+\frac{1}{4}}\rho^se^{-\frac{\rho}{2}}\sqrt{\Gamma(2s+n+1/2)\Gamma(n+1)}\times\\
&\times\sum_{j=0}^n\frac{\left(\frac{-\zeta^*}{\left(1-|\zeta|^2\right)}\right)^j}{\Gamma(j+1)\Gamma(2s+\frac{1}{2}+n-j)}\sum_{p=0}^\infty\frac{\zeta^p}{p!}\frac{\Gamma(n-j+p+1)}{\Gamma(n-j+1)}L^{2s-\frac{1}{2}}_{n-j+p}(\rho).
\end{align}
The procedure to obtain the explicit form of these number states is
explained in references \cite{Nos1}. It consists in use the sums (48.7.6) and (48.7.8) of reference \cite{ERH}.
Thus, the explicit form of the $SU(1,1)$ Perelomov number coherent states of the pseudoharmonic oscillator is
\begin{align}
\Psi_{PO}=&\left(\frac{1-|\zeta|^2}{(1-\zeta)^2}\right)^{s+\frac{1}{4}}\rho^se^{-\frac{\rho}{2}}e^{\left(\frac{\rho\zeta}{\rho-1}\right)}\sqrt{\frac{\Gamma(n+1)}{\Gamma(2s+n+1/2)}}\left(-\zeta^*\right)^n(1-\sigma)^n\times\\
&\times L^{2s-\frac{1}{2}}_n\left(\frac{\rho\sigma}{(1-\zeta)(\sigma-1)}\right),
\end{align}
where
\begin{equation}
\sigma=\left(\frac{1-|\zeta|^2}{\zeta^*(1-\zeta)}\right).
\end{equation}

\section{$SU(2)$ Perelomov number coherent states}

In what follows, the results for the $su(2)$ Lie algebra are
obtained in a similar way to those for the $su(1,1)$ Lie algebra.
The $su(2)$ Lie algebra is spanned by the generators $J_{+}$,
$J_{-}$ and $J_{0}$, which satisfy the commutation relations
\cite{Vourdas}
\begin{eqnarray}
[J_{0},J_{\pm}]=\pm J{\pm},\quad\quad [J_{+},J_{-}]=2J_{0}.\label{comsu2}
\end{eqnarray}
The action of these operators on the Fock space states
$\{|j,\mu\rangle, -j\leq\mu\leq j\}$  is
\begin{equation}
J_{+}|j,\mu\rangle=\sqrt{(j-\mu)(j+\mu+1)}|j,\mu+1\rangle,\label{j+n}
\end{equation}
\begin{equation}
J_{-}|j,\mu\rangle=\sqrt{(j+\mu)(j-\mu+1)}|j,\mu-1\rangle,\label{j-n}
\end{equation}
\begin{equation}
J_{0}|j,\mu\rangle=\mu|j,\mu\rangle.\label{j0n}
\end{equation}
The displacement operator $D(\xi)$ is
\begin{equation}
D(\xi)=\exp(\xi J_{+}-\xi^{*}J_{-}).
\end{equation}

By means of Gaussian decomposition, the normal form of this operator is
\begin{equation}
D(\xi)=\exp(\zeta J_{+})\exp(\eta J_{0})\exp(-\zeta^*J_{-}),\label{normalsu2}
\end{equation}
where $\zeta=-\tan(\frac{1}{2}\theta)e^{-i\varphi}$ and $\eta=-2\ln \cos |\xi|=\ln(1+|\zeta|^2)$ \cite{Perellibro}.

The $SU(2)$ Perelomov coherent states, $|\zeta\rangle=D(\xi)|j,-j\rangle$ are
given by \cite{Perellibro}
\begin{equation}
|\zeta\rangle=\sum_{\mu=-j}^j\left[\frac{(2j)!}{(j+\mu)!(j-\mu)!}\right]^{\frac{1}{2}}(1+|\zeta|^2)^{-j}\zeta^{j+\mu}|j,\mu\rangle.\label{PCSsu2}
\end{equation}

In a similar way to the definition (\ref{defPCNS}), the Perelomov
number coherent states for the $su(2)$ algebra are defined as the
action of the displacement operator $D(\xi)$ on any state
$|j,\mu\rangle$, instead to the lower state $|j,-j\rangle$ of the
Fock space. Thus,
\begin{equation}
|\zeta,j,\mu\rangle=D(\xi)|j,\mu\rangle=\exp(\zeta J_{+})\exp(\eta
J_{0})\exp(-\zeta^* J_{-})|j,\mu\rangle\label{defPCNSsu2}
\end{equation}
where we have used the normal form of the displacement operator, equation
(\ref{normalsu2}).

Therefore, the Perelomov number coherent states of the $su(2)$ algebra in the Fock space are given by \cite{Nos2}
\begin{eqnarray}
|\zeta,j,\mu\rangle&=&\sum_{s=0}^{j-\mu+n}\frac{\zeta^s}{s!}\sum_{n=0}^{\mu+j}\frac{(-\zeta^*)^n}{n!}e^{\eta(\mu-n)}
\frac{\Gamma(j-\mu+n+1)}{\Gamma(j+\mu-n+1)}\nonumber\\
&&\times\left[\frac{\Gamma(j+\mu+1)\Gamma(j+\mu-n+s+1)}{\Gamma(j-\mu+1)\Gamma(j-\mu+n-s+1)}\right]^{1/2}|j,\mu-n+s\rangle.\label{PNCSSU2}
\end{eqnarray}
The $SU(2)$ standard coherent states of equation (\ref{PCSsu2}) are recovered by setting $\mu=-j$ in the last equation.

The similarity transformation of the $su(2)$ Lie algebra generators
are computed by using of the Baker-Campbell-Hausdorff identity and
equations (\ref{comsu2}). They are
\begin{equation}
D^{\dag}(\xi)J_{+}D(\xi)=-\frac{\xi^{*}}{|\xi|}\delta
J_{0}+\epsilon\left(J_{+}+\frac{\xi^{*}}{\xi}J_{-}\right)+J_{+},\label{simiJ+}
\end{equation}
\begin{equation}
D^{\dag}(\xi)J_{-}D(\xi)=-\frac{\xi}{|\xi|}\delta
J_{0}+\epsilon\left(J_{-}+\frac{\xi}{\xi^{*}}J_{+}\right)+J_{-},\label{simiJ-}
\end{equation}
\begin{equation}
D^{\dag}(\xi)J_{0}D(\xi)=(2\epsilon+1)
J_{0}+\frac{\delta\xi}{2|\xi|}J_{+}+\frac{\delta
\xi^*}{2|\xi|}J_{-},\label{simiJ0}
\end{equation}
where $\delta=\sin(2|\xi|)$ and
$\epsilon=\frac{1}{2}\left[\cos(2|\xi|)-1\right]$.

From equations (\ref{simiJ+})-(\ref{simiJ0}), the expected values of the group generators $J_{\pm}, J_0$ in the Perelomov number coherent states
are
\begin{equation}
\langle \zeta,j,\mu|J_{+}|\zeta,j,\mu\rangle=\frac{\xi^*}{|\xi|}\mu\sin(2|\xi|),
\end{equation}
\begin{equation}
\langle \zeta,j,\mu|J_{-}|\zeta,j,\mu\rangle=\frac{\xi}{|\xi|}\mu\sin(2|\xi|),
\end{equation}
\begin{equation}
\langle \zeta,j,\mu|J_{0}|\zeta,j,\mu\rangle=\mu\cosh(2|\xi|).
\end{equation}

\subsection{Schrödinger uncertainty relationship}

The $X$ and $Y$ operators for the $su(2)$ algebra ladder operators are defined as \cite{Chinos}
\begin{eqnarray}
X\equiv J_{+}+J_{-},\quad\quad Y\equiv i(J_{+}-J_{-}).
\end{eqnarray}

The quadratic deviations product for the $X$ and $Y$ operators in the
$SU(2)$ Perelomov number coherent states is
\begin{equation}
(\Delta X)^{2}_{n}(\Delta Y)^{2}_{n}=4(j+j^2-\mu^2)^2\left\{(\epsilon^2+\epsilon)^2\left[4-\left(\frac{\xi^*}{\xi}+\frac{\xi}{\xi^*}\right)^2\right]+4(\epsilon^2+\epsilon)+1\right\}.\label{prodxynsu2}
\end{equation}

If we use equation (\ref{defPCNSsu2}) and the similarity transformation
method, the expectation values $\langle F \rangle$ and
$\langle C \rangle$ in a $SU(2)$ number coherent states are given by
\begin{equation}
\langle\zeta,j,\mu|F|\zeta,j,\mu\rangle_{n}=2i(j+j^2-\mu^2)(\epsilon^2+\epsilon)\left(\frac{\xi^{*}}{\xi}-
\frac{\xi}{\xi^{*}}\right),\label{fnsu2}
\end{equation}
\begin{equation}
\langle\zeta,j,\mu|C|\zeta,j,\mu\rangle_{n}=-4\mu(2\epsilon+1)\label{cnsu2}.
\end{equation}
By substituting the results of equations (\ref{prodxynsu2}),
(\ref{fnsu2}) and (\ref{cnsu2}) into equation (\ref{schro}) we
conclude, again, that the $SU(2)$ Perelomov number coherent states are not of minimum
uncertainty, accordingly to the Schrödinger uncertainty
relationship. However, likewise the $SU(1,1)$ standard coherent states, the $SU(2)$
standard coherent states satisfy the equality in equation (\ref{schro}). Therefore, the only
states which minimize the Schrödinger uncertainty relationship are
those obtained by applying the displacement operator $D(\xi)$ on the
lowest normalized state.

\subsection{Time evolution of the $SU(2)$ Perelomov number coherent states}

As for the case of the $su(1,1)$ algebra, we will suppose that the
Hamiltonian is proportional to the group generator $J_{0}$. Hence
\begin{equation}
U(t)=e^{-i\gamma J_{0}t/\hbar}.
\end{equation}
This implies that the time evolution of the $su(2)$ algebra ladder
operators $J_{\pm}$ are
\begin{equation}
J_{+}=U^{\dag}(t)J_{+}U(t)=J_{+}e^{i\gamma t/\hbar},\label{j+t}
\end{equation}
\begin{equation}
J_{-}=U^{\dag}(t)J_{-}U(t)=J_{-}e^{-i\gamma t/\hbar}.\label{j-t}
\end{equation}

Thus, by using equation (\ref{defPCNSsu2}), the time evolution of
the $SU(2)$ number coherent states $|\zeta(t),j,\mu\rangle$ are given by
\begin{equation}
|\zeta(t),j,\mu\rangle=U(t)|\zeta\rangle=U(t)D(\xi)U^{\dag}(t)U(t)|j,\mu\rangle.\label{zeta,t,su2}
\end{equation}
From equation (\ref{j0n}), (\ref{j+t}) and (\ref{j-t}), and the
definitions  $\xi(t)\equiv\xi e^{i\gamma t/\hbar}$ and
$\zeta(t)=\zeta e^{i\gamma t/\hbar}$, we can show that the time
dependent $SU(2)$ Perelomov number coherent states are given by
\begin{equation}
|\zeta(t),j,\mu\rangle=e^{-i\gamma\mu t/\hbar}e^{\zeta(-t)
J_{+}}e^{\eta J_{0}}e^{-\zeta(-t)^*
J_{-}}|j,\mu\rangle.\label{NPCSt}
\end{equation}
Therefore, the time evolution of these states is obtained by adding the phase $e^{-i\gamma\mu t/\hbar}$ and substituting $\zeta\rightarrow\zeta(-t)$ and $\zeta^*\rightarrow\zeta(-t)^*$ into equation (\ref{PNCSSU2}). The expression of equation (\ref{PNCSt}) generalizes the $SU(2)$ Perelomov coherent states, which are recovered by setting $t=0$ and $\mu=-j$.

\subsection{$SU(2)$ coherent states for the two-dimensional harmonic oscillator}

The time-independent Hamiltonian of the two-dimensional harmonic oscillator is
\begin{equation}
H=a^{\dag}a+b^{\dag}b+1,
\end{equation}
where the operators $(a, a^{\dag})$ and $(b, b^{\dag})$ satisfy the bosonic algebra
\begin{equation}
[a,a^{\dag}]=[b,b^{\dag}]=1, \quad\quad[a,b^{\dag}]=[a,b]=0.
\end{equation}
The Jordan-Schwinger realization of the $su(2)$ algebra is
\begin{equation}
J_0=\frac{1}{2}\left(a^{\dag}a-b^{\dag}b\right), \quad J_+=a^{\dag}b, \quad J_-=b^{\dag}a,\label{jordan}
\end{equation}
For this realization the Casimir operator and the number operator $N$ commute with all
the generators of the $su(2)$ algebra. The number operator $N$ is defined as
\begin{equation}
N=a^{\dag}a+b^{\dag}b.
\end{equation}
The eigenfunctions of this Hamiltonian $H$ are
\begin{equation}
\langle\rho,\phi|N,m\rangle=\psi_{N,m}(\rho,\phi)=\frac{1}{\sqrt{\pi}}e^{im\phi}(-1)^{\frac{N-m}{2}}\sqrt{\frac{2\left(\frac{N-m}{2}\right)!}{\left(\frac{N+m}{2}\right)!}}
\rho^{m}L_{\frac{1}{2}(N-m)}^{m}(\rho^2)e^{-\frac{1}{2}\rho^2}.\label{function}
\end{equation}
The action creation and annihilation operators on the basis $|N,m\rangle$ is given by \cite{wallace}
\begin{eqnarray}
a|N,m\rangle=\sqrt{\frac{1}{2}(N+m)}|N-1,m-1\rangle,\quad a^{\dag}|N,m\rangle=\sqrt{\frac{1}{2}(N+m)+1}|N+1,m+1\rangle,\label{acta}
\end{eqnarray}
\begin{eqnarray}
b|N,m\rangle=\sqrt{\frac{1}{2}(N-m)}|N-1,m+1\rangle,\quad b^{\dag}|N,m\rangle=\sqrt{\frac{1}{2}(N-m)+1}|N+1,m-1\rangle.\label{actb}
\end{eqnarray}
From these equations and the definition of the $su(2)$ generators of equation (\ref{jordan}) we can obtain the relationships between the group numbers $j, \mu$
and the quantum numbers $N, m$. Thus, from equations (\ref{j0n}) and (\ref{j+n}) we deduce $\mu=m/2$, $j=N/2$.

In order to obtain the $SU(2)$ Perelomov coherent states of the two-dimensional harmonic oscillator $\Psi_{HO}$ we must substitute the eigenstates (\ref{function}) into equation (\ref{PCSsu2}).
By making the change of variable $s=j+\mu$ in equation (\ref{PCSsu2}) we obtain
\begin{equation}
\Psi_{HO}=\langle\rho|\zeta\rangle=\sqrt{\frac{2(2j)!}{\pi}}\frac{e^{-\frac{1}{2}\rho^2}}{(1+|\zeta|^2)^{-j}}\sum_{s=0}^{2j}\frac{\zeta^s (-1)^{2j-s}e^{2i(s-j)\phi}\rho^{2(s-j)}}{s!}L_{2j-s}^{2s-2j}(\rho^2).
\end{equation}
This sum can be performed by using the equation (48.19.5) of reference \cite{ERH}
\begin{equation}
\sum_{k=0}^{n}\frac{n!(-1)^k p^k}{(n-k)!}L_k^{n-2k}(x)=p^{\frac{n}{2}}H_n\left[\frac{1}{2}(1+px)p^{-1/2}\right].
\end{equation}
Then, by identifying $k=2j-s$, $n-2k=2s-2j$ we finally obtain the closed form of the $SU(2)$ coherent states for the two-dimensional harmonic oscillator
\begin{equation}
\Psi_{HO}=\sqrt{\frac{2}{N!\pi}}\frac{e^{-\frac{1}{2}\rho^2}\zeta^{N/2}}{(1+|\zeta|^2)^{N/2}}H_N\left[\frac{\sqrt{\tau}}{2}\left(1+\frac{\rho^2}{\tau}\right)\right],
\end{equation}
where
\begin{equation}
\tau=\frac{1}{\zeta\rho^2 e^{2i\phi}}. 
\end{equation}
It is important to note that we were not able to obtain explicitly the $SU(2)$ Perelomov number coherent states. The main problem is that for the $SU(2)$ case we must perform two finite series,
instead of one finite and one infinite of the $SU(1,1)$ case.

\section{Concluding remarks}

We have studied some properties of the Perelomov number coherent states for the $su(1,1)$ and
$su(2)$ Lie algebras. We introduced the position and
momentum-like operators and showed that the Schrödinger uncertainty
relationship is minimized only for the standard Perelomov coherent
states.

We apply our results to calculate the explicit form of the $SU(1,1)$ Perelomov
number coherent states of the pseudoharmonic oscillator. For the two-dimensional harmonic oscillator,
we were able to calculate the explicit form of the standard $SU(2)$ Perelomov coherent states.

Besides the Perelomov number coherent states are not of minimum uncertainty, they
have been applied to solve some important quantum systems as the parametric amplifier
and two coupled oscillators.

\section*{Acknowledgments}
This work was partially supported by SNI-M\'exico, COFAA-IPN,
EDI-IPN, EDD-IPN, SIP-IPN project number $20161727$.

\end{document}